\documentstyle[12pt]{article}

\begin{document}
\begin{titlepage}
\begin{center}
November 24, 1997     \hfill    LBNL-41106 \\

\vskip .5in

{\large \bf Mermin's Suggestion and the Nature of Bohr's Action-at-a-Distance
Influence}

\footnote{This work was supported by the Director, Office of Energy 
Research, Office of High Energy and Nuclear Physics, Division of High 
Energy Physics of the U.S. Department of Energy under Contract 
DE-AC03-76SF00098.}
\vskip .30in
Henry P. Stapp\\
{\em Lawrence Berkeley National Laboratory\\
      University of California\\
    Berkeley, California 94720}
\end{center}

\vskip .3in

\begin{abstract}

Mermin  suggests  comparing my recent  proof of  quantum  nonlocality to Bohr's
reply to Einstein, Podolsky, and Rosen. Doing so leads naturally to the insight
that the nonlocal  influence deduced from the  analysis of the Hardy experiment
is the same as the nonlocal influence deduced by Bohr, and used by him to block
the  application of the  criterion of  physical  reality  proposed by Einstein,
Podolsky,   and  Rosen.  However,  the  greater    sophistication of  the Hardy
experiment, as  contrasted to the  experiment considered  by Bohr and the three
authors,  exposes more clearly  than before  the nature of  this influence, and
thereby strengthens Bohr's position. 

\end{abstract}
\vskip 0.5in 
\begin{center}
Full postscript file available from\\
http://physweb.lbl.gov/theorygroup/papers/
\end{center}
\end{titlepage}

\renewcommand{\thepage}{\roman{page}}
\setcounter{page}{2}
\mbox{ }

\vskip 1in

\begin{center}
{\bf Disclaimer}
\end{center}

\vskip .2in

\begin{scriptsize}
\begin{quotation}
This document was prepared as an account of work sponsored by the United
States Government. While this document is believed to contain correct 
 information, neither the United States Government nor any agency
thereof, nor The Regents of the University of California, nor any of their
employees, makes any warranty, express or implied, or assumes any legal
liability or responsibility for the accuracy, completeness, or usefulness
of any information, apparatus, product, or process disclosed, or represents
that its use would not infringe privately owned rights.  Reference herein
to any specific commercial products process, or service by its trade name,
trademark, manufacturer, or otherwise, does not necessarily constitute or
imply its endorsement, recommendation, or favoring by the United States
Government or any agency thereof, or The Regents of the University of
California.  The views and opinions of authors expressed herein do not
necessarily state or reflect those of the United States Government or any
agency thereof or The Regents of the University of California and shall
not be used for advertising or product endorsement purposes.
\end{quotation}
\end{scriptsize}

\vskip 2in

\begin{center}
\begin{small}
{\it Lawrence Berkeley Laboratory is an equal opportunity employer.}
\end{small}
\end{center}

\newpage
\renewcommand{\thepage}{\arabic{page}}
\setcounter{page}{1}

Mermin's article$^1$ fingers a key  point. I had taken an even-handed approach,
and made no  attempt to single  out which of  my explicitly  stated assumptions
fails. At least one of them must be invalid, since their conjunction leads to a
logical contradiction$^2$. Mermin argues that LOC2 fails.

Mermin then suggests  linking the putative failure  of LOC2 to Bohr's reply$^3$
to the paper of Einstein,  Podolsky, and Rosen  (EPR)$^4$. That is an excellent
idea. The  meaning and  validity of  Bohr's reply  has been much  debated. Karl
Popper$^5$   claims that   Einstein, not  Bohr, won  that  famous  battle. John
Bell$^6$ likewise  has questioned the  rationality of  Bohr's argument. Linking
Bohr's   argument to  the  presumed  failure  of LOC2  turns  out to  be indeed
illuminating: it  supplies the very  element that critics  of that argument had
deemed missing.

I reject Mermin's other  suggestion that there as a  `hidden gap' or `essential
ambiguity'  in my  argument: the  assumption  LOC2 in  question,  is explicitly
stated and technically  unambiguous. The pertinent  question is rather what its
failure would mean: in what sense, if any, would its violation entail some sort
of faster-than-light influence. I answer that key question now.

The logical form of LOC2 is this:

We have proved (Mermin grants), under the condition that L2 is performed
in the left-hand region, the truth of the follow statement:

(S): If performing R2 were to give +, then performing R1, instead, must give -.

Statement (S) was proved, as the initial proviso indicates, under the condition
that  L2 was   performed in  the  faraway  space-time  region  L,  which can be
considered to lie later in time than  the space-time region R within which the
possible experiments R1 and R2 and their possible results are confined.

That is, the direct  proof of (S) would fail if L1  were to be performed at the
later time rather than L2.

But LOC2 claims  that statement (S),  which refers explicitly  only to possible
experiments  and results  confined to R,  cannot be true if  someone at a later
time freely chooses to do one thing  but false if that person freely chooses to
do something else instead.

Mermin claims that I `overlooked' essentially the fact that the direct proof of
(S) fails if L2 is  not performed.  Exactly the opposite  is true: I introduced
LOC2 precisely to get  around that problem. The  pertinent issue is not whether
the direct  proof  continues to  hold under  condition  L1---it  certainly does
not---: it is rather whether a nontrivial dependence of the truth of (S) upon a
future free choice means that there is some sort of backward-in-time influence,
even though the statement (S) is a counterfactual assertion.

The point, as I see it, is that if L2  is chosen, then it is provably true that
in any instance for which R2 gives +, R1 would have given -: e.g., every time a
`color'  measurement  gives ``red'',  not  ``green'', a  `firmness' measurement
would necessarily have given ``hard'', not ``soft''. Although the proof of this
connection  depends  on the  later free  choice being  L2, the  conclusion is a
connection  between  two earlier   properties: redness  entails  hardness. Even
though both properties  cannot be simultaneously  determined empirically, there
is an  invariable   relationship  between what  happens  under two  alternative
conditions.  This  means  that  Nature's choices  under  these two  alternative
earlier conditions must  be correlated in a certain  way, provided L2 is freely
chosen at the later time, but---if LOC2 is violated---cannot correlated in this
way if the  later free  choice is L1.  But this  means that if  LOC2 were to be
violated then there must  be some sort of  backward-in-time influence: Nature's
choices under alternative possible  earlier experimental conditions necessarily
must be  constrained in  one way if  the later free  choice is  L2, but in some
different  way if the later free choice is L1.

I formulated my locality  condition in terms of the  dependence of the truth of
statements pertaining to  `possible measurements  and their possible observable
outcomes that are all  localized in a space-time  region that lies earlier than
some time T': the  locality condition  was that the truth  of such statements 
cannot depend upon free choices made by experimenters after time T.

Mermin asserts  that the  failure of this  locality condition  ``does not mean,
however, that a choice  made in the future (whether  to perform L1 or L2 on the
the left) can influence events in the  present.'' However, the term ``influence
events'' contains an essential ambiguity. What certainly is true is that is the
faster-that-light    influence  entailed  by a  failure of  LOC2 is  not of the
simplest  kind: it  involves not just  differences  of what  happens in the one
actually occurring  experimental situation, under different faraway conditions,
but  rather   differences in  the  contraints  imposed  upon  Nature's  earlier
selections under  the different  conditions that we are  free to set up faraway
and later. 

This situation, in which there is a  subtle action-at-a-distance influence, but
no simple direct one, certainly brings  to mind Bohr's basic claim in his reply
to EPR:

[T]here is... no question of a mechanical disturbance of the system under
investigation...[but] there is essentially the question of an influence
on the very conditions which define the possible types of predictions regarding
the future behavior of the system.

In  both  cases  there is a   denial of  any  (proof of a)  direct  mechanical
disturbance  of  events in the  single  actually  occurring  situation, but an
affirmation of the existence of an influence of some kind.

So I follow Mermin's suggestion that  an examination of this similarity between
what  Bohr  claims and  what a   failure of  LOC2  entails  ``might  offer some
illumination to those  have difficulty  understanding Bohr's reply to Einstein,
Podolsky, and Rosen (EPR)''.

The meaning of Bohr's argument has been much debated. Mermin cites Plotnitsky's
book$^7$ for a ``thoughtful critique of Bell's statements about Bohr's views''.
Plotnitsky roundly condemns Bell as completely failing to understand Bohr. Bell
himself admits to not  understanding Bohr's  argument, but with the implication
that Bohr's argument does not make  sense. Plotnitsky cites some key phrases of
Bohr that Bell  did not  mention, namely ``the  wording of the  above mentioned
criterion  of  physical  reality  proposed  by  Einstein,  Podolsky,  and Rosen
contains an ambiguity as regards the  meaning of the expression `without in any
way disturbing a system' '', and [at  the end of a passage cited above] ``these
conditions  constitute an inherent  element of any  phenomena to which the term
`physical reality' can be  applied.'' Plotnitsky  suggests that the reason that
Bell does not understand Bohr's argument is perhaps that he refuses to read the
full argument that gives meaning to these important phrases. I am confident
that Bell, a thorough scientist, did examine Bohr's full argument carefully 
before publically criticizing it.

What Bohr's full  argument shows is that if we take  the point of view that the
quantum formalism is a procedure for  making, on the basis of knowledge that we
acquire by performing possible  measurements, predictions about the outcomes of
our later possible  observations, then  performing one of  the earlier possible
measurements may exclude the  possibility of performing an alternative possible
one. Applied to the particular example  that EPR considered, this consideration
was shown  to entail  that if  one sets up  the system  so that  $q_1 +q_2$ and
$p_1-p_2$ are both  well defined, as  EPR specify, then  one can measure either
$q_1$ or $p_1$, but not both, and  hence become able to predict either $q_2$ or
$p_2$ but  not both.  Thus, from  Bohr's  perspective, in which  the meaning of
``physical  reality'' is  tied to our  acquiring  the knowledge  needed to make
predictions about it, measuring $q_1$  does disturb the other system because it
produces ``an influence on the very  conditions which define the possible types
of predictions  regarding future  behavior of the [other]  system.'' Thus if we
measure $q_1$ then we cannot make a prediction about $p_2$.
 
By this argument Bohr disputes the key  EPR claim that performing a measurement
on one of two   correlated---but currently   non-interacting---systems does not
``disturb'' the other.

So the point of  Bohr's argument is to  assert that within  his knowledge-based
and prediction-oriented  Copenhagen framework there  IS an action-at-a-distance
influence, and the existence of this  action-at-a-distance influence blocks the
EPR (implicit)  claim that there is  none, thereby  blocking application of the
EPR criterion of reality

How does Bohr's claim that there IS an  important  action-at-distance influence
relate to my similar claim?

The EPR-Bohr debate centered on the EPR criterion for physical reality:

``If, without in any way disturbing a system, we can predict with certainty the 
value of a physical quantity then there exists an element of physical reality
corresponding to that physical quantity''.

The central  question in  judging the adequacy  of Bohr's  reply is whether the
faster-than-light  influence that he claims exists,  within his knowledge-based
framework of thinking,  can be both sufficiently  real to block the application
of the EPR criterion of physical reality, yet sufficiently unreal to produce no
mechanical  disturbance. That is the  problem that  troubled Bell, and probably
everyone else who is troubled by Bohr's argument

A main objective of my  work was to rid the  argument of these ``reality''
questions that have led into a philosophical quadmire of interminable disputes.

In Mermin's section ``What's wrong?'' he says that the step from line 5 to line
6 of my proof is wrong. But the  failure of this step is exactly the claim that
LOC2, as it is applied, is not valid. That is just what I am trying to show, or
in any case is  sufficient.  The crucial  question, however,  is whether such a
failure of LOC2 entails some sort of faster-than-light influence. I have given,
above, a direct argument that it does.

Mermin gets involved in questions of definitions and meanings, as did Bohr. But
Bohr had to  involve himself with such  matters because he  was confronted by a
characterization of ``physical reality'' that was basically alien to what arises
from his  own  knowledge-based  approach.  Hence he was  forced,  in effect, to
redefine  this key  term to  bring it  into line  with his  own  philosophy. My
approach is designed to circumvent these definitional issues.

Specifically,   Mermin questions  whether  the fact  that  statement (S) refers
explicitly only to  possible  measurement and possible  results confined to the
right-hand region R really justifies making the step from line 5 to line 6? His
point  is that  the  original  proof  of (S)  depends  on the  fact  that L2 is
performed in region L, and so he would like therefore simply to stick with that
result and hence allow the truth of  (S) to depend on the future choice between
L1 and L2.  But I claim  that my  argument given  above shows  that denying the
validity  of  LOC2 in  this  way does  entail  some  sort of   backward-in-time
influence  of the  later free  choice on  Nature's  earlier  selection process:
certain rigid  connections between  what these earlier  selections can be under
alternative possible conditions in region R must exist if L2 is performed in L,
but must fail to hold if L1 is performed there. 

This result pertaining  to LOC2, carried over to  the EPR-Bohr case, would mean
that in  that case  connections  between  Nature's  choices of the  outcomes of
measuring  $q_2$ or  alternatively $p_2$  cannot be assumed  not to depend upon
whether $q_1$ or  $p_1$ is measured.  But this means that  one cannot pass from
the fact that one can measure either $q_1$ or $p_1$, and hence determine either
$q_2$ or  $p_2$, to the  conclusion that both  results are  simultaneously well
defined: one cannot assume that all connections between outcomes of alternative
possible   measurments that  could be  performed  in one  spacetime  region are
independent of what  someone will choose to do  faraway and later. This result,
which is what the  analysis of the Hardy experiment  shows, gives solid support
to Bohr's  position as opposed  to EPR. So I  think that the  faster-than-light
influence that  I have  established is  basically in line with  Bohr's reply to
EPR, and indeed that the  faster-than-light influence that Bohr deduced from an
analysis of possible knowledge-based predictions is essentially the same as the
one that I  deduced  from an  analysis of  the Hardy  experiment.  However, the
sophistication   of the  Hardy   experiment, in  comparison  to the  simple EPR
experiment, allows the nature of this  influence to be exposed now more clearly
than before. This identification of my  faster-than-light influence with Bohr's
entails that this  faster-than-light influence is,  as Bohr claimed but critics
doubted, logically sufficienty to block the application of the EPR criterion of
physical reality.

I  have adopted, throughout,  the    straight-forward   meanings  of  logical
propositions, and  not tried to evaded  drawing the  anti-intuitive conclusions
that logic demands by adopting some  rule that capriciously asserts well-formed
propositions  to be   ``meaningless'',  rather than  merely true  or false, and
provable or  unprovable,  when they  have a  perfectly clear  meaning. Nor do I
think that Bohr's policy  was one of the evasion of  logic. Quite the opposite!
He did not  seek to denigrate  or deny an   action-at-a-distance influence that
logic called for  by curtailing the  power of logic.  Rather he pushed logic to
the limit  to show  that  rational analysis  demanded  an  action-at-a-distance
influence, and did not shirk from accepting that conclusion even though it flew
in the  face of  common  sense  classical   intuition. He  used the  `essential
ambiguity'  that he  identified, which  concerned  what he  viewed as an overly
restrictive view of the meaning of  ``without in any way disturbing a system'',
to  enlarge  the  meaning  of   ``disturb'',  not  to  curtail  its  meaning by
arbitrarily limiting the scope of logical reasoning. 

The  major  weakness in  Bohr's  position,  however, was  that the  action at a
distance   whose  existence  he  claimed  seemed, to  many  critics,  to be too
unphysical  to really  do the  job. This  action seemed  to exist  only in some
nebulous realm, that might be  `defined' to be physical reality, but that seems
too  abstract  and  ghostly to  really  impact on  physical  reality.  Thus the
Hardy-based   analysis  fortifies  Bohr's  position  at its  weakest  point, by
allowing the needed  action at a distance to be  deduced more directly from the
unquestioned structure of  the predictions of the  theory than from a debatable
philosophical commitment concerning the nature of the scientific endeavour, and
a radical re-interpretation of the meaning of physical reality.
\newpage
{\bf References}

1. N. David Mermin, ``Nonlocal character of quantum theory?'', 
   American Journal of Physics, preceeding article.

2. Henry P. Stapp,  ``Nonlocal character of quantum theory'', 
   American Journal of Physics, {\bf 65}, 300-304 (1997).
   For a shorter version with only two locality conditions, LOC1 and LOC2,
   see ``Quantum Ontology and Mind-Matter Synthesis'' on my website:
   http://www-physics.lbl.gov/$\sim$stapp/stappfiles.html

3. N Bohr, ``Can quantum-mechanical description of physical reality be
   considered complete?'', Phys. Rev. {\bf 48}, 696-702 (1935).     

4. Albert Einstein, Boris Podolsky, and Nathan Rosen,
   ``Can quantum-mechanical description of physical reality be
   considered complete?'', Phys. Rev. {\bf 47}, 777-780 (1935).     

5. Karl Popper, ``Quantum mechanics without `The Observer' '', in 
   {\it Quantum Theory and Reality}, ed. Mario Bunge, Springer-Verlag, 
   Berlin, (1967). 

6. John S. Bell, ``Bertlemann's socks and the nature of reality'',
   Appendix 1, in {\it Speakable and Unspeakable in Quantum Mechanics},
   Cambridge (1987), 155-156.

7. Arkady Plotnitsky, {\it Complementarity}, Duke University Press (1994),
    172-190.

\end{document}